\documentclass{article}
\usepackage{spconf,amsmath,graphicx,hyperref}

\title{Thinking While Listening: \\ Simple Test Time Scaling For Audio Classification}
\name{Prateek Verma \qquad Mert Pilanci}
\address{Department of Electrical Engineering \\
Stanford University \\
Stanford, CA 94305, USA}
\begin{document}
%
\maketitle
\begin{abstract}
We propose a framework that enables neural models to “think while listening” to everyday sounds, thereby enhancing audio classification performance. Motivated by recent advances in the reasoning capabilities of large language models, we address two central questions: (i) how can thinking be incorporated into existing audio classification pipelines to enable reasoning in the category space and improve performance, and (ii) can a new architecture be designed from the ground up to support both thinking and test-time scaling? We demonstrate that in both settings, our models exhibit improved classification accuracy. Leveraging test-time scaling, we observe consistent gains as the number of sampled traces increases. Furthermore, we evaluate two open-source reasoning models, GPT-OSS-20B and Qwen3-14B, showing that while such models are capable of zero-shot reasoning, a lightweight approach—retraining only the embedding matrix of a frozen, smaller model like GPT-2 can surpass the performance of billion-parameter text-based reasoning models.
\end{abstract}
\section{Introduction and Related Work}
\label{sec:intro}
In recent years, Large Language Models (LLMs) have transformed artificial intelligence, tackling challenges once considered intractable, such as solving olympiad problems \cite{trinh2024solving}. Their influence now extends beyond text, shaping natural language processing \cite{brown2020language}, acoustic modeling via tokens \cite{borsos2023audiolm}, raw audio generation \cite{verma2021generative}, computer vision \cite{yan2021videogpt}, and robotics \cite{brohan2023rt2}. A common paradigm is to discretize the modality into tokens and train a GPT-style model for next-token prediction, with reconstructions performed in the target domain if necessary from the discrete tokens.
Beyond training, strategies such as test-time scaling \cite{muennighoff2025s1,brown2024large} have further advanced performance. For example, \cite{brown2024large} demonstrated that verifying multiple sampled outputs of a LLM and selecting the best, optimal for a chosen criteria, improves the performance with the number of times we sample the output. In contrast with this, our work leverages LLMs to consume sequences of patch-level predictions (“reasoning trace”) and directly output categories, without having explicit verification modules. Test-time scaling provides a powerful inference-time paradigm: the model and input remain fixed, while performance improves as the model “thinks longer” by generating and refining alternative solutions. In this paper, we adapt this framework to the domain of audio classification. Audio classification has historically followed advances in other domains, employing architectures like CNNs \cite{hershey2017cnn} and generative models \cite{verma2020framework}. Performance gains have traditionally been achieved by scaling model parameters or dataset sizes \cite{ellis1999size}. Specific to audio, another option has been to build a better front end keeping other parts of the model fixed  \cite{ravanelli2018speaker,zeghidour2021leaf,verma2023content} More recently, test-time scaling has emerged as a viable alternative, with methods like multi-level augmentations showing significant metric boosts in audio-captioning \cite{Kim2022Exploring}. Extending this paradigm, we demonstrate that classification performance can be improved at test time even with a fixed model and input. We propose a method where, instead of naive input transformations, we aggregate evidence from stochastic audio patch predictions via reasoning model. We build a category trace that reflects the classifier's evolving hypothesis over time. Our LLM-based aggregation is inspired by Chain-of-Thought (CoT) prompting \cite{wei2022chain}, which elicits intermediate reasoning steps in LLMs to improve performance, even in zero-shot setup \cite{kojima2022large}. This is enhanced through self-consistency over multiple reasoning chains \cite{lu2022frozen}. While recent work explored CoT for audio from raw tokens \cite{ma2025audio}, our approach complements this by constructing the reasoning chain from the outputs of a perceptual model. Here, the ``steps" of reasoning correspond to evidence from different audio patches rather than textual exemplars. In summary, our contribution is a ``thinking-while-listening" pipeline: at test time, we sample patch-level predictions from a frozen audio classifier to build a reasoning trace. A frozen LLM then leverages this trace to refine the final classification.\begin{figure*}[t]
    \centering
    \includegraphics[width=0.9\textwidth,height=8.5cm]{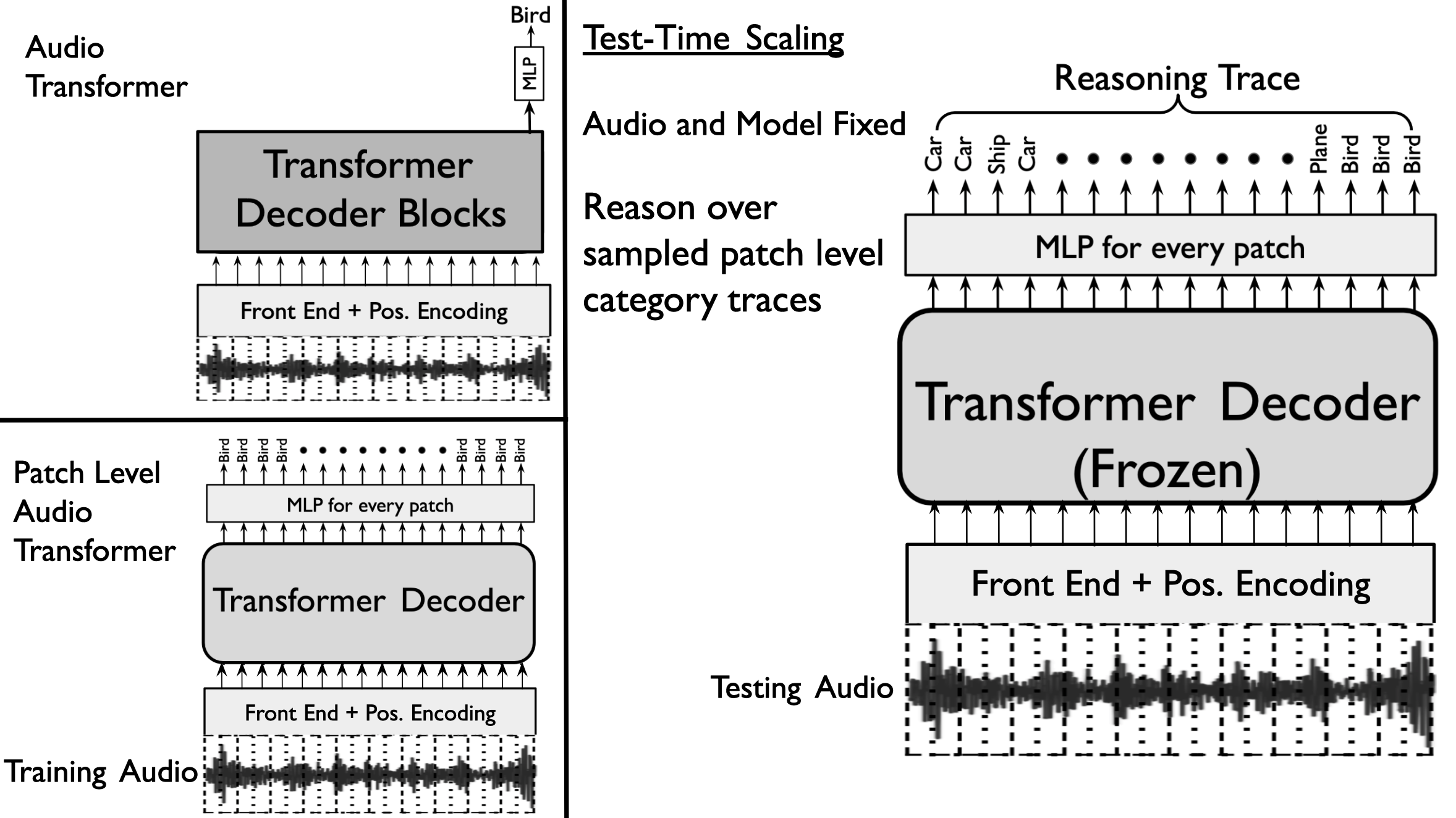}
    \caption{Description of our proposed method. During training, we allow the model to causally predict categories for every patch to get the patch-level category output. The trained model is frozen during inference, and the audio is fixed. We sample from the posterior probability category distribution of each patch multiple times (which we define as the length of the sampling trace of a patch) to get a reasoning trace. The trace is then used to understand the category using a frozen LLM reasoning model like GPT-OSS 20B or a frozen GPT-2 model with a new embedding matrix to aggregate and give accurate predictions for the audio.}
    \vspace{-0.5cm}
    \label{fig:tts_esc50}
\end{figure*}
\vspace{-0.25cm}
\section{Methodology}
\label{sec:pagestyle}
We extend pre-trained audio models such as Audio Spectrogram Transformers (AST) \cite{gong21b_interspeech} and YAMNet \cite{YAMNet} to generate stepwise reasoning traces, and also explore new models designed for this purpose. Motivated by advances in NLP, we adopt strategies from LLAMA \cite{touvron2023llama} and GPT \cite{radford2019language}, where intermediate tokens (e.g., \textless think\textgreater, \textless wait\textgreater) elicit structured reasoning. Large-scale pretraining enables such tokens to invoke latent reasoning behaviors, which are leveraged through prompting after initial predictions. These reasoning prompts guide frozen models to refine outputs iteratively. Our goal is to construct analogous reasoning traces in audio tasks to enhance classification, with reasoning aided from text LLMs.

\subsection{Reasoning Traces From Existing Classification Model} We show how audio understanding models developed as simple classification pipelines can be converted into models that can think and give us reasoning traces. In the context of this paper, the reasoning trace is a collection of category labels of what the model thinks the audio is until the waveform it has heard. The reasoning trace is a string of categories of length equal to the number of patches we break down the audio into casually, times the length of the sampling trace per patch, and is explained in the next subsection. For experiments with fixed 5s audio inputs in case of ESC-50 dataset, we vary the sampling trace length by segmenting audio into 500ms patches. In AST, predictions are generated causally after each new patch, concatenating them to increase the context for every output. As expected, initial predictions say the first 500ms are unreliable, yet performance improves steadily with longer context. This process yields a reasoning trace: at each patch, the sampling trace length corresponds to the number of times output probabilities are collected. Thus, a 5s audio produces $10 \times T$ category predictions, where $T$ is the sampling trace length, each interleaved with posterior confidence. The full reasoning trace length is therefore $20 \times T$. Since the model is pretrained on AudioSet, categories are drawn from the AudioSet ontology \cite{gemmeke2017audio}. While these may differ from the downstream dataset, semantic overlap allows the reasoning module to adapt effectively. Since models like YAMNet \cite{YAMNet} and convolutional architectures such as VGG or ResNet \cite{hershey2017cnn} typically output a single embedding for a fixed audio input, adapting them to provide a reasoning trace is non-trivial. Similar to AST-style setups, for YAMNet we process audio in 500ms increments, generating predictions at each step. The convolutional encoder remains fixed, while predictions evolve with context length. Longer contexts yield more confident and accurate classifications. Interestingly, even short segments yield informative and reasonable predictions, often capturing fine-grained cues absent when processing full context as model tries hard to predict the best category in constrainted context.
\subsection{Models That Can Think While Listening} 
Building on the initial results, we extend existing models such as Audio Transformers \cite{verma2021audio} to generate reasoning traces at test time, enabling a single forward pass over frozen audio and model inputs. As shown in Fig. 1, the baseline employs a single MLP head on the final Transformer decoder embedding, with AUC scores reported accordingly. Instead of training with a single target vector, category probabilities are predicted per patch during training using a sigmoid layer, with mean-squared error loss minimization. Since each 1s segment of FSD-50K may contain multiple categories, we modify the sampling process. Each MLP head which outputs a 200-dimensional vector is normalized to make it a probability distribution for patch-wise sampling. Both models are trained for 300 epochs with a learning rate of 1e-3, decaying to 1e-6.The front-end employs 64 filters and six layers of embedding dimension 64, similar to Audio Transformer \cite{verma2021audio}.

\subsection{Test Time Scaling Over Reasoning Traces} For both YAMNet (convolutional) and AST (transformer-based) models, audio is segmented into 500\,ms chunks, and performance is evaluated as a function of sampling length. The per-chunk predictions form a reasoning trace that is aggregated into a final prediction, with majority voting as the simplest aggregator. In S1 \cite{muennighoff2025s1}, multiple outputs were combined using special tokens (e.g. \textless think\textgreater) to simulate step-wise reasoning, while for scaling methods using verifier aggregation is carried by sampling multiple candidates choose best optimized using verifier. In our setup, reasoning traces are generated causally: at each step, the model predicts likely categories conditioned on the audio observed. Candidates are sampled from the posterior distribution, where $T$ defines the number of samples per patch, yielding a reasoning sequence of length $2P \times T$ for $P$ patches, with confidence. This sequence is fed to reasoning models such as GPT-OSS 20B \cite{GPT-OSS-20B} and Qwen3 14B \cite{Qwen-14B}, guided by a structured prompt \footnote{Prompt to GPT-OSS 20B and Qwen3 14B

``We take an audio wavform of 5 seconds and divide it into 10 patches each of 500ms. For each of the patch we sample multiple times and list the categories sampled from the distribution. For the entire audio waveform, can you predict which category the sound belongs to. For predicting the best category DO NOT COUNT or take the MEAN of the predicted categories. Rather reason through the category traces from patch 0 to 9 in a sequential manner. Reason and take into account what constitues a particular sound, what sub-atoms of a sound an audio is made of and draw the correlation from the category labels predicted to what best the sound patch and the entire trace  progression would be. Take into account the confidence scores for each patch in the range of 0-100 with 100 being very confident and 0 being not at all confident for each of the patches. Here are the details of the audio file:  The number of times each patch is sampled: 32. 

CURRENT PATCH 0  -- Categories for patch are: list of categories/conf.

CURRENT PATCH 1  -- Categories for patch are: list of categories/conf.

So on ....

LIST OF CATEGORIES GIVEN

From the list please pick only one category most likely to be the audio

} 
to give the output category. This is a zero-shot setup, as models like S1, do not introduce additional parameters and keep the LLM decoder backbone frozen. We feed this trace to a frozen GPT-2 base model, re-training only the embedding matrix while keeping the weights and positional embeddings fixed, following the approach of Audio PALM \cite{rubenstein2023audiopalm}. The new embedding matrix vocabulary matches the number of problem categories, with 10 extra confidence tokens representing confidence scores (0–1) in 10 buckets, and the total output categories in the dataset of interest (50 for ESC-50, 200 for FSD-50K). Category tokens are interleaved with confidence tokens to form the reasoning trace, yielding $2PT$ tokens as input to GPT-2 (embedding size 768) across all experiments. This setup activates existing LLM connections while keeping the backbone fixed. Unlike \cite{rubenstein2023audiopalm}, we pass the last token prediction through an MLP classification head with sigmoid activation \cite{hershey2017cnn} for multi-label outputs. Merely tuning the embedding matrix with frozen GPT-2 weights improves performance over open-source reasoning models (GPT-OSS4, Qwen3) using complex prompts. The performance improves with stronger reasoning models and increased length of posterior sampling for audio models.
\section{Dataset}
\vspace{-0.5cm}
\label{sec:format}
We evaluate our framework on two widely used public audio classification datasets with available audio samples: ESC-50 \cite{piczak2015esc50} and FSD-50K \cite{fonseca2020fsd50k}, covering both single- and multi-label classification setups. The ESC-50 dataset consists of 50 audio categories with a total of 2,000 clips, each 5 seconds long and associated with a single label corresponding to the contents present in the audio. For this setup, we use backbone models built on pretrained networks, namely YAMNet and AST, and keep their weights frozen throughout training and inference. In contrast, FSD-50K involves predicting 200 categories from 1-second audio segments, where multiple labels may be present per clip. We adopt a consistent experimental protocol across both datasets: for ESC-50, we report top-1 accuracy, while for FSD-50K we evaluate performance using AUC for multi-label predictions. In both cases, we compare baseline model performance against results obtained with our test-time scaling approach, highlighting the improvements achieved by sampling techniques at inference. 

\begin{figure}[t]
    \centering
    \includegraphics[width=0.9\linewidth]{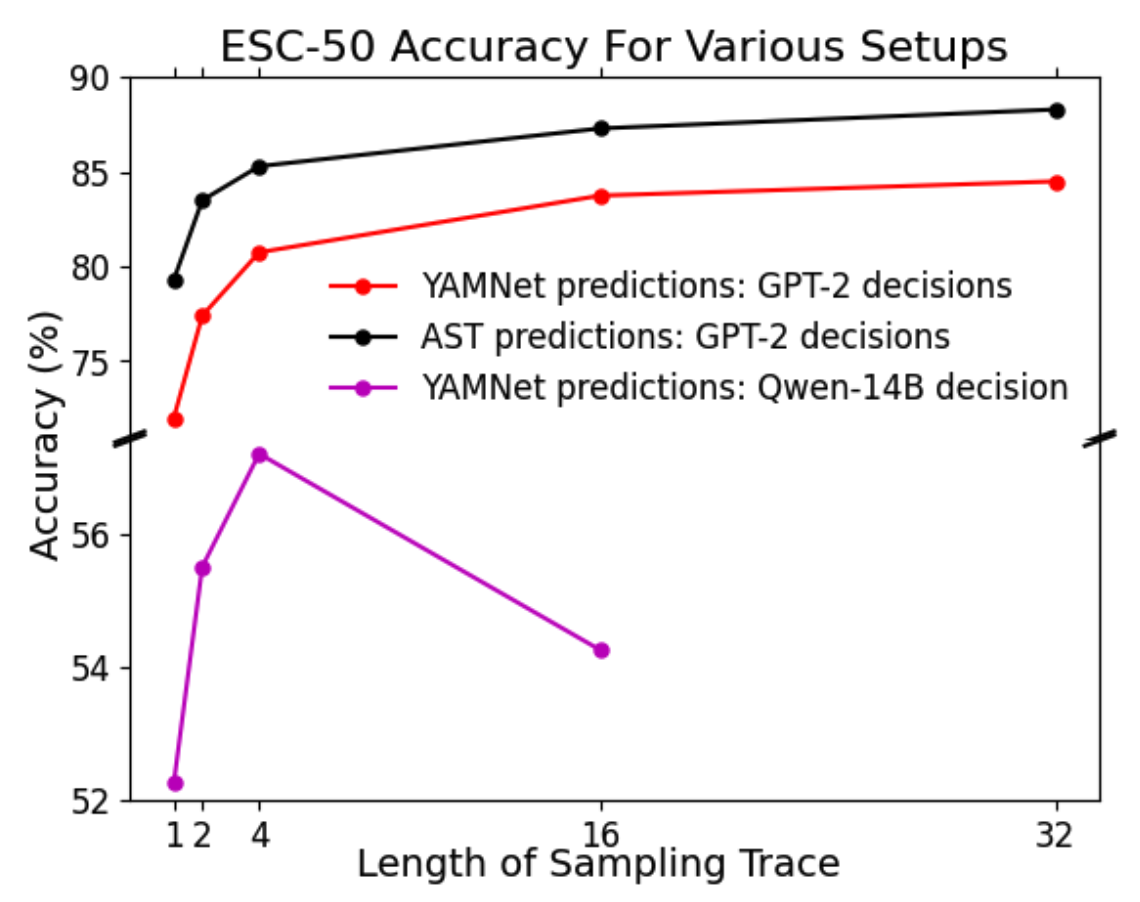}
    \caption{Results for Test Time Scaling For ESC50 dataset for frozen YAMNet and AST as a function of length of sampling trace, using GPT-2 and Qwen3-14B for category prediction. }
    \label{fig:result_esc50}
\vspace{-0.3cm}
\end{figure}
\section{Experiments and Results} 
\label{sec:typestyle}
\vspace{-0.3cm}
Our experiments highlight two consistent sources of improvement: (i) stronger pretrained reasoning models and (ii) longer patch-level sampling traces. We first evaluate frozen audio backbones on ESC-50, specifically YAMNet and AST trained on AudioSet with baseline as 84\% on ESC-50 with model frozen. Full fine-tuning AST on ESC-50 yields 88.8\% \cite{gong21b_interspeech}. We can see that a frozen model can be pushed to a full-tuned performance with that of our proposed test-time scaling paradigm. As shown in Fig. 2, top-1 accuracy increases with trace length, with patch-level predictions fed into pretrained reasoning models including GPT-OSS 20B, Qwen3-14B, and frozen GPT-2 with a newly trained embedding matrix \cite{radford2019language}. Despite frozen weights, AST consistently outperforms YAMNet, and both backbones benefit from test-time scaling. Table 2 indicates that GPT-OSS and Qwen do not surpass frozen GPT-2 with re-trained embedding matrix for this task. However, GPT-OSS designed for chain-of-thought reasoning—outperforms Qwen in our setup showing gains in reasoning performance carry forward to our task. Notably, all experiments keep audio backbones and transformer blocks of reasoning models fixed. Test-time performance improves with more posterior samples giving longer reasoning traces through repeated patch-level predictions. This establishes a robust test-time scaling paradigm, in contrast to standard audio classification pipelines, such as AST and YAMNet, which output a single probability vector. Adjusting temperature provides a minor gain (0.2\%), insufficient to justify a large-scale study under the current setup.\begin{figure}[t]
    \centering
    \includegraphics[width=\linewidth]{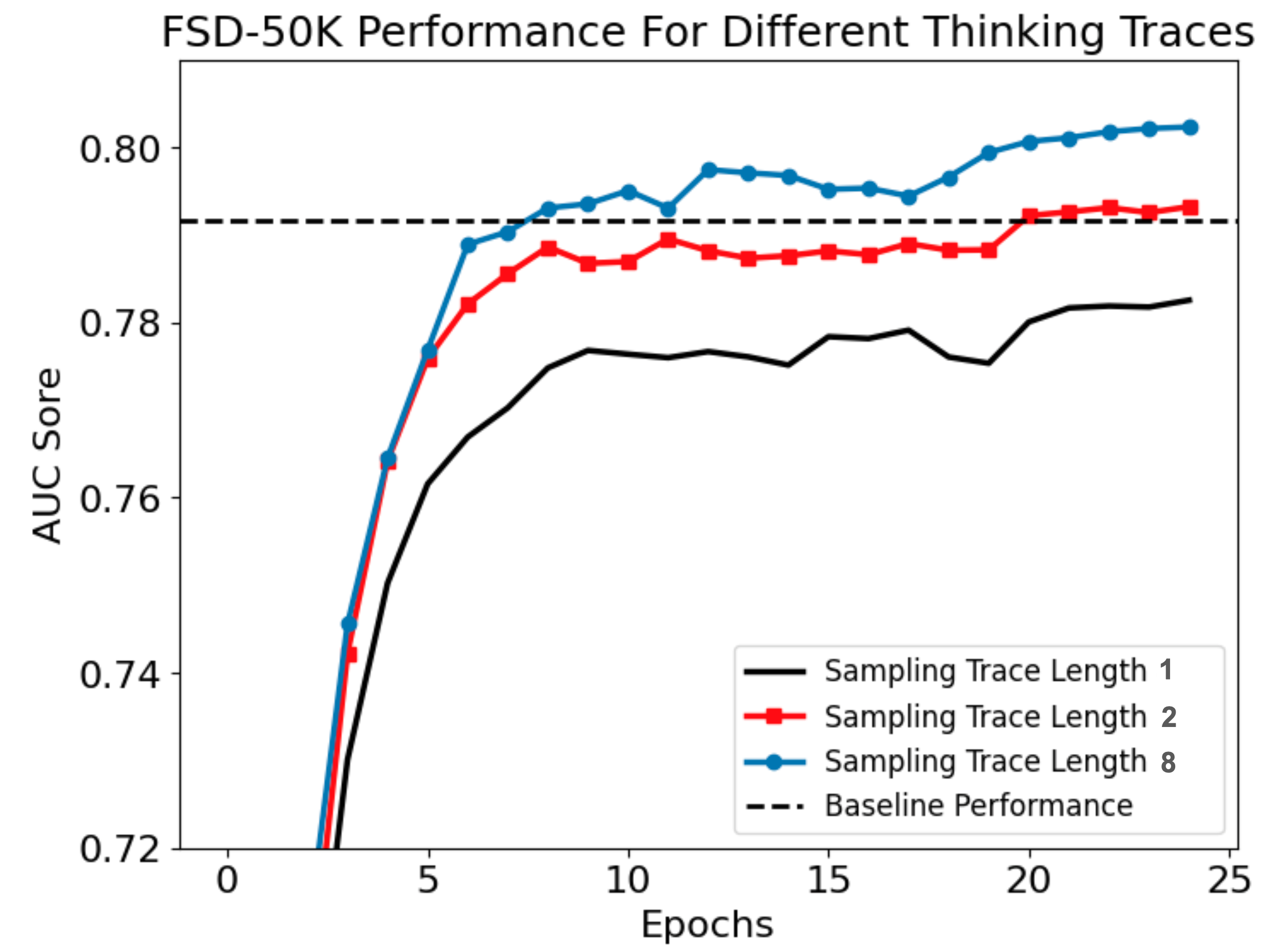}
    \caption{Results on 1s chunks on FSD050K dataset for different length of sampling traces. We reason through the sampling trace with a frozen GPT-2 backbone, with trained embedding matrix with for baseline Audio Transformer, and backbone. }
    \label{fig:result_fsd50k}
\end{figure} For our proposed architecture, we extend the framework to patch-level multi-label prediction by dividing each second into 40 patches similar to Audio Transformer baseline \cite{verma2021audio}. As shown in Fig. 3, shorter sampling traces do not improve over the baseline Audio Transformer AUC on 1-second inputs, likely due to information loss when sampling over single categories. Increasing the trace length (1 → 2 → 8 samples per patch) consistently boosts performance. When sampling at the patch level eight times per patch—our method surpasses the baseline. Across experiments, performance improves with longer traces and stronger reasoning models. Even weaker audio backbones, such as YAMNet, benefit from test-time scaling, demonstrating that model improvements are achievable regardless of initial backbone strength. Better architectures gives accurate reasoning traces, which in turn enhance test-time behavior.

\begin{table}[t]
\centering
\caption{ESC-50 accuracy with different sampling length for YAMNet backbone and zero shot text reasoning models.}
\label{tab:model_temperature}
\begin{tabular}{c|cccc}
\hline
\textbf{Model} & \multicolumn{4}{c}{\textbf{Sampling length/ output prediction}} \\
\cline{2-5}
 & \textbf{1} & \textbf{2} & \textbf{4} & \textbf{16} \\
\hline
GPT-OSS 20B & 53.5 & 58.75 & 57.6 & \textbf{61.25} \\
Qwen3 14B  & 52.3  & 55.5 &  57.2 & 54.25 \\\hline
\end{tabular}
\end{table}
\begin{table}[t]
\centering
\caption{ESC-50 accuracy with different temperatures/sampling trace lengths for AST backbone frozen, making predictions in 500ms increments using Frozen GPT-2.}
\label{tab:temperature}
\begin{tabular}{c|c|ccccc}
\hline
\textbf{Temp} & \textbf{Model} & \multicolumn{5}{c}{\textbf{Sampling length / op prediction}} \\
\cline{3-7}
 &  & \textbf{1} & \textbf{2} & \textbf{4} & \textbf{16} & \textbf{32} \\
\hline
1.0 & YAMNet & 72.0 & 77.4 & 80.8 & 83.8 & 84.5  \\\hline
1.0 & AST & \multicolumn{4}{c}{{\textit{Full Model Finetune} \cite{gong21b_interspeech}}} & \textbf{88.8} \\
1.0 & AST & 79.3 & 83.5 & 86.3 & 87.3 & \textbf{88.3}  \\
1.2 & AST & 76.8 & 84.8 & 85.3 & 87.0 & 87.0 \\
1.5 & AST & 72.5 & 80.5 & 82.8 & 86.5 & \underline{\textbf{88.5}} \\
2.0 & AST & 53.5 & 65.3 & 77.3 & 84.8 & 83.8 \\
\hline
\end{tabular}
\end{table}
\section{Conclusion and Future Work}
\label{sec:conclusion} We present a simple approach for incorporating test-time scaling in audio classification, applicable to both existing pretrained models and a newly designed architecture that supports sampling of “reasoning traces” conditioned on an input audio signal. In addition, we improve baseline model by allowing multiple category predictions at each patch, which yields consistent gains as the sampling trace length increases. We provide ablation studies that integrate text-based reasoning and demonstrate that a frozen GPT-2, used for this task solely by retraining its embedding matrix, further improves performance. The proposed methods can be easily integrated into standard audio or image classification pipelines. Traditionally, advances in model performance have been attributed to increased scale—either in the number of parameters or the size of training data. In contrast, our results highlight a complementary direction: performance can be further scaled by reasoning over patch-level category traces during inference, while keeping both the backbone weights and input audio fixed. We observe this effect across two standard audio benchmarks, underscoring the generality of our approach. Our method is sufficiently general to enable test-time performance scaling for advancing a wide range of applications.

\bibliographystyle{IEEEbib}
\bibliography{refs}

\end{document}